\documentclass{osa-article}
\usepackage[Symbol]{upgreek}
\usepackage{amsmath}
\usepackage{amssymb}
\usepackage{mathrsfs}
\usepackage{graphicx}
\usepackage{dcolumn}
\usepackage{bm}
\usepackage{hyperref}
\usepackage{float}

\journal{osac}

\articletype{Research Article}

\begin{document}

\title{Two-dimensional-controlled high-order modes and vortex beams from an intracavity mode converter laser}

\author{Jing Pan\authormark{1,2}, Yijie Shen\authormark{3}, Zhensong Wan\authormark{1,2}, Xing Fu\authormark{1,2}, Hengkang Zhang\authormark{1,2}, and Qiang Liu\authormark{1,2,*}}

\address{\authormark{1}Key Laboratory of Photonic Control Technology (Tsinghua University), Ministry of Education, Beijing 100084, China\\
\authormark{2}State Key Laboratory of Precision Measurement Technology and Instruments, Department of Precision Instrument, Tsinghua University, Beijing 100084, China\\
\authormark{3}Optoelectronics Research Centre, University of Southampton, Southampton SO17 1BJ, United Kingdom}

\email{\authormark{*}Corresponding author: qiangliu@tsinghua.edu.cn} 



\begin{abstract}
We present a novel scheme of structured light laser with an astigmatic mode converter (AMC) as intracavity element, first enabling the generation of Hermite-Gaussian (HG) modes with fully controlled two-dimensional (2D) indices $(m,n)$ and vortex beams carrying orbital angular momentum (OAM) directly from cavity. The 2D tunability was realized by controlling the off-axis displacements of both pump and intracavity AMC. The output HG$_{m,n}$ beam could be externally converted into OAM beam with 2D tunable radial and azimuthal indices $(p,\ell)$. With the certain parameter control, OAM$_{0,\pm1}$ vortex beam also could be directly generated from the cavity. Our setup provides a compact and concise structured light source. It has great potential in extending various applications of optical tweezers, communications, and nonlinearity.
\end{abstract}


Vortex beams carrying orbital angular momentum (OAM) were uesd in particle manipulation, high-security encryption, communications, and nonlinear optics~\cite{shen2019optical,fang2019orbital,wang2012terabit,rego2019generation}, with tunable azimuthal index $\ell$ related to OAM as a researching focus. However, as another independent index, radial index $\textit{p}$ was barely noted, which could be related to the radial momentum of light and applied in quantum entanglement~\cite{chen2019realization}, encoding information~\cite{karimi2014radial}, and mode sorting~\cite{gu2018gouy}. For extending more usages of OAM beams, it is desired to expand tunable mode range from one dimension (index $\ell$ only) to two dimensions, with both radial and azimuthal indices $(p,\ell)$, which has profound impacts on the applications mentioned above. On the other hand, it is also desired to generate beams carrying OAMs directly from cavity, which can simplify the setup of OAM beams~\cite{forbes2019structured}. However, based on the classic and simple way to generate vortex beams by using cylindrical lens as astigmatic mode converter (AMC) to transform HG$_{m,n}$ into LG$_{p,\ell}$ beams and using off-axis pumping in solid-state laser to generate required HG modes~\cite{laabs1996excitation,chen1997generation,shen2018wavelength}, only HG modes from cavity, with the order tunable in only one dimension, and the converted OAM beams with untunable radial index ($\textit{p}=0$)~\cite{shen2018wavelength} could be generated. There also existed some methods to generate two-dimensional (2D) high-order modes directly from a laser, such as introducing 2D gain~\cite{kong2012generation,shen2018vortex}. However, these methods prevented the mode indices from being freely tuned, and the vortex and non-vortex modes being freely switched. Therefore, it urgently requires a new method to freely control the 2D indices of HG modes and switchable vortex modes at the source. 
  
In this letter, we proposed a new structured light laser which could controllably generate HG modes with 2D-tunable indices directly from a laser and switchable vortex beams. We elaborately put an AMC in an off-axis-pumped laser cavity. The output HG modes could have tunable 2D indices controlled by the displacements of pump and AMC. With another AMC as the external converter of vortex beams, modes carrying OAMs were obtained with continually and independently tunable azimuthal and radial indices $(p,\ell)$. As another important function, the structured laser cavity could also directly generate the OAM beams by certain parameter control. The whole setup was concise, compact and cost-saving. The method can enrich many applications and offer a unique insight into the formation of structured modes in laser cavities. 

\begin{figure}[htbp]
\centering
\includegraphics[width=8cm]{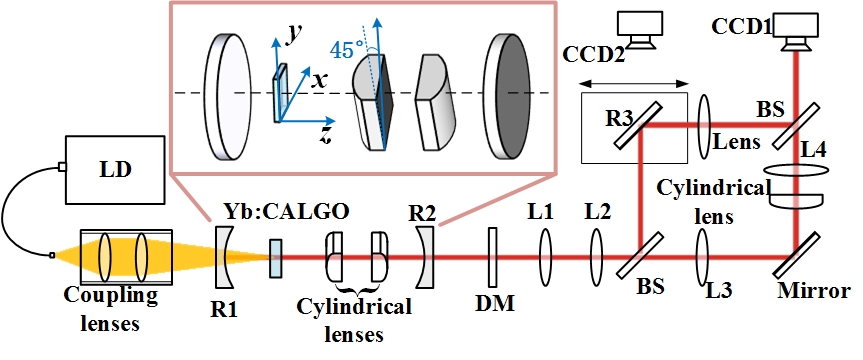}
\caption{The experimental setup with the insert showing the orientation of cylindrical lenses.}
\label{fig:1}
\end{figure}

The experimental setup is shown in Fig.~\ref{fig:1},  with an AMC in the cavity. The pumping source was a 976~nm fiber-coupled laser diode (LD, Han's TCS, highest power: 110 W) with 105~$\mu$m fiber core and 0.22~NA numerical aperture. With two 976~nm antireflective (AR) coated lenses [focal lengths ($F$): 30~mm and 60~mm respectively] separated by 20~mm, the pumping beam was focused into the gain medium, an a-cut Yb:CALGO. A concave dichroic mirror [R1, AR at 979~nm and high-reflective (HR) at 1040-1080~nm, radius of curvature: 1200~mm] was 90~mm away from the AR coated lenses. The Yb:CALGO ($4\times4\times2$ mm$^3$, 5~at.\%, AR at 940-1080~nm) wrapped by a cooling heat sink of $18^{\circ}C$ was set 30~mm away from R1, and the output coupler (R2, transmittance was 1\% at 1030-1080~nm, radius of curvature: 300~mm) was placed on the other side of Yb:CALGO. As the control elements to stimulate a new tuning dimension of modes, two cylindrical lenses ($F=$25~mm both) separated by 36~mm inclined to $45^{\circ}$ were set between Yb:CALGO and R2, 21.3~mm away from Yb:CALGO and 7~mm away from R2. One cylindrical lens was used for controlling the direction of HG modes in one dimension along the cylindrical lens's generatrix, and the other cylindrical lens was used for controlling the order in another dimension with its off-axis displacement. A dichroic mirror (DM: reflectance is 99\% at 979 nm and transmittance is 90\% at 1050-1080 nm) was set outside the cavity to filter off excess pumping light. The introduction of two cylindrical lenses in the cavity led to astigmatic 2D HG beams. HG beams went through a lenses group and a cylindrical lens outside the cavity to form OAM beams. Being 60~mm away from the DM, a lenses group including lenses L1 ($F=$25~mm), L2 ($F=$30~mm), and L3 ($F=$100~mm) was placed consecutively, with the separation of 100~mm and 128~mm respectively. A cylindrical lens ($F=$25~mm) placed vertically was set 100~mm away from L3 to achieve astigmatic conversion. Lens L4 ($F=$100~mm) was placed 80~mm away from the cylindrical lens, and images were collected by charge coupled device (CCD1, Spiricon, M2-200s) 100~mm away from L4. To check the topological charges of OAM beams, we established an interference system. A beamsplitter (BS: transmittance and reflectance both are 50\% at 1064 nm) was inserted $45^{\circ}$ from the transmission direction between L2 and L3 to lead off part of light as the reference light. Along the direction of reference light, one mirror (R3) set parallel with the BS was used to fold the reference light to form a Mach-Zehnder cavity. A lens with the focal length of 100~mm was set after R3 for beam expansion. OAM beams and the reference light converged through another BS with the same parameters as the former one to form the interference pattern, recorded by CCD1. When R3 was removed, CCD2 behind R3 recorded patterns of HG modes.

Cylindrical lenses introduced astigmatism, due to their asymmetry in two directions, into the output modes. Based on the analysis of astigmatism in the cavity~\cite{nienhuis2007structure,habraken2008orbital,habraken2009rotationally}, the axes of two cylindrical lenses in our cavity were parallel to each other, and therefore we considered the astigmatism in our cavity as a simple one, and treated two dimensions separately~\cite{nienhuis2007structure}.
The beam parameters can be characterized by ABCD matrix.
In the cavity, parameter $1\diagup q$ is given as:
\begin{equation}
\frac{1}{q}=\frac{D-A}{2B}-\textit{i}\frac{1}{\vert B\vert}\sqrt{1-\left(\frac{A+D}{2} \right)^{2} }
\label{eq:1}
\end{equation}
The imaginary part of \textit{q} is $q_{0} $, and the real part of \textit{q} is the distance \textit{z} away from the equivalent beam waist, shown as:
\begin{equation}
\begin{split}
&q_{0}=\frac{i\frac{1}{\vert B\vert}\sqrt{1-\left(\frac{A+D}{2} \right)^{2} }}{\left(\frac{D-A}{2B}\right)^{2}+\frac{1}{\vert B\vert^{2}}\left[ 1-\left(\frac{A+D}{2} \right)^{2} \right] }\\
&z=\frac{\frac{D-A}{2B} }{\left(\frac{D-A}{2B}\right)^{2}+\frac{1}{\vert B\vert^{2}}\left[ 1-\left(\frac{A+D}{2} \right)^{2} \right]  }
\end{split}
\label{eq:2}
\end{equation}
Expressed with parameter $q_{0}$, the radius of beam waist is $w_{0}=\sqrt{-\textit{i}q_{0} \lambda\diagup\pi}$ and the Rayleigh length $z_{R}$ is $z_{R}=-\textit{i}q_{0}$. Based on the experiment, due to the modes’ astigmatism relating to the off-axis displacement of R1 and the limitation of cylindrical lens to the pumping light, the ratio of $w_{0_{x}}$ and $ w_{0_{y}}$ was led in.

ABCD matrices of the cavity perpendicular to and parallel with the generatrices of cylindrical lenses are given as Eq.~(\ref{eq:3}), respectively, substituted into Eq.~(\ref{eq:1}) and Eq.~(\ref{eq:2}).

\begin{equation}
\begin{split}
\textbf{X}=&\textbf{M}(d_1)\textbf{N}\left(-\frac{1}{f_1}\right)\textbf{M}(d_2)\textbf{N}\left(-\frac{1}{f_2}\right)\textbf{M}(d_3)\textbf{N}\left(-\frac{2}{R_2}\right)\textbf{M}(d_3)\\&\textbf{N}\left(-\frac{1}{f_2}\right)\textbf{M}(d_2)\textbf{N}\left(-\frac{1}{f_1}\right)\textbf{M}(d_1)\textbf{N}\left(-\frac{2}{R_1}\right)\\
\textbf{Y}=&\textbf{M}(d_1)\textbf{M}(d_2)\textbf{M}(d_3)\textbf{N}\left(-\frac{2}{R_2}\right)\textbf{M}(d_3)\textbf{M}(d_2)\textbf{M}(d_1)\textbf{N}\left(-\frac{2}{R_1}\right)
\end{split}
\label{eq:3}
\end{equation}
where $\textbf{M}(x)$$=$$\begin{bmatrix}
1&x\\
0&1
\end{bmatrix}$, $\textbf{N}(x)$$=$$\begin{bmatrix}
1&0\\
x&1
\end{bmatrix}$, $\textbf{X}$$=$$\begin{bmatrix}
A_{x}&B_{x}\\
C_{x}&D_{x}
\end{bmatrix}$, $\textbf{Y}$$=$$\begin{bmatrix}
A_{y}&B_{y}\\
C_{y}&D_{y}
\end{bmatrix}$. $d_{1}$, $d_{2}$, $d_{3}$ are the distances between R1 and the first cylindrical lens, the first cylindrical lens and the second cylindrical lens, the second cylindrical lens and R2. $R_{1}$ and $R_{2}$ are the radii of curvature of R1 and R2, and $f_{1}$ and $f_{2}$ are the focal lengths of two cylindrical lenses, respectively.

Astigmatic HG mode generated from the cavity is given by:
\begin{equation}
\begin{split}
\Psi_{m,n}^{(\rm HG)}\left(x,y\right)={(2^{m+n-1}\pi m!n!)}^{-1/2}\psi_m\left(x\right)\psi_n\left(y\right)\exp\left(ikz\right)
\end{split}
\label{eq:4}
\end{equation}
where HG function yields ($\xi=x$, $t=m$; $\xi=y$, $t=n$)
\begin{equation}
\psi_t\left(\xi\right)=\frac{e^{-\frac{\xi^{2}}{\omega_{\xi}^{2}(z_{\xi})}}}{\sqrt{\omega_\xi}}H_{t}\left[ \frac{\sqrt{2}\xi}{\omega_{\xi}(z_{\xi})}\right]\\
e^{ik\frac{\xi^{2}}{2\sqrt{R_{x}(z_{x})R_{y}(z_{y})}}}
e^{-\textit{i}\left( t+\frac{1}{2}\right)\tan ^{-1}\left(\frac{z_{\xi}}{z_{R_\xi}} \right)},
\label{eq:5}
\end{equation}
where $\omega\left(z\right)=\omega_{0}\sqrt{1+\left( {z}/{z_{R}}\right)^{2} },R\left(z \right)=z\left(1+{z_{R}^{2}}/{z^{2}} \right) $


ABCD matrices outside the cavity, substituted into formulas about $\omega$ and $\textit{q}$ in Eq.~(\ref{eq:8}), are given as:
\begin{equation}
\begin{split}
\textbf{A}=&\textbf{M}(d_6)\textbf{N}\left(-\frac{1}{f_5}\right)\textbf{M}(d_5)\textbf{N}\left(-\frac{1}{f_4}\right)\textbf{M}(d_4)\textbf{N}\left(-\frac{1}{f_3}\right)\textbf{M}(d_3)\\&\textbf{N}\left(-\frac{1}{f_2}\right)\textbf{M}(d_2)\textbf{N}\left(-\frac{1}{f_1}\right)\textbf{M}(d_1+L-z_{x})\\
\textbf{B}=&\textbf{M}(d_6)\textbf{N}\left(-\frac{1}{f_5}\right)\textbf{M}(d_5)\textbf{N}\left(-\frac{1}{f_4}\right)\textbf{M}(d_4)\textbf{N}\left(-\frac{1}{f_3}\right)\textbf{M}(d_3)\\&\textbf{N}\left(-\frac{1}{f_2}\right)\textbf{M}(d_2)\textbf{N}\left(-\frac{1}{f_1}\right)\textbf{M}(d_1+L-z_{y})
\end{split}
\label{eq:6}
\end{equation}
where $\textbf{M}(x)$$=$$\begin{bmatrix}
1&x\\
0&1
\end{bmatrix}$, $\textbf{N}(x)$$=$$\begin{bmatrix}
1&0\\
x&1
\end{bmatrix}$, $\textbf{A}$$=$$\begin{bmatrix}
A_{1x}&B_{1x}\\
C_{1x}&D_{1x}
\end{bmatrix}$, $\textbf{B}$$=$$\begin{bmatrix}
A_{1y}&B_{1y}\\
C_{1y}&D_{1y}
\end{bmatrix}$. $L$ is the length of the cavity, and $d_{1}$ to $d_{6}$ are distances between R2 and L1, L1 and L2, L2 and L3, L3 and the cylindrical lens, the cylindrical lens and L4, L4 and CCD1. $f_{1}$ to $f_{3}$, and $f_{5}$ are focal lengths of L1 to L4, and $f_{4}$ is the focal length of the cylindrical lens. HG modes could be written as~\cite{huang2018large}:
\begin{equation}
\begin{split}
\Psi_{m,n}^{(\rm HG)}\left(x,y\right)={(2^{m+n-1}\pi m!n!)}^{-1/2}\psi_{m}\left(x\right)\psi_{n}\left(y\right)
\end{split}
\label{eq:7}
\end{equation}
where HG function yields ($\xi=x$, $t=m$; $\xi=y$, $t=n$)
\begin{equation}
\psi_t\left(\xi\right)=\frac{H_{t}\left[ \frac{\sqrt{2}\xi}{\omega_{\xi}(z_{\xi})}\right]}{\sqrt{\omega_{0_\xi}}} 
\left( \frac{1}{A_{1\xi}+\frac{B_{1\xi}}{q_{0_\xi}}}\right)^{t+\frac{1}{2}}
\left(\frac{\omega_{\xi}}{\omega_{0_\xi}}\right)^{t}\exp\left(-i\frac{\pi \xi^{2}}{\lambda q_{\xi}}\right)
\label{eq:8}
\end{equation}
where $\omega=\omega_{0}\sqrt{\left(A_{1}+\frac{B_{1}}{q_{0}} \right)^2+i\frac{ 2B_{1}\lambda\left( A_{1}+\frac{B_{1}}{q_{0}}\right)}{ \pi\omega_{0}^{2}  }}$, $q=\frac{A_{1}q_{0}+B_{1}}{C_{1}q_{0}+D_{1}}$. After astigmatic conversion outside the cavity, modes carrying OAMs are written as:
\begin{equation}
\Psi^{(\rm OAM)}\left(x,y\right)=\sum_{s=0}^{N}d_{s-\frac{N}{2},n-\frac{N}{2}}^{\frac{N}{2}}\left( \theta\right)\times \Psi_{s,N-s}^{\left(\rm HG\right)}\left(x,y\right)
\label{eq:9}
\end{equation}
where $N=n+m$, and

\begin{align}
\nonumber
&d_{s-\frac{N}{2},n-\frac{N}{2}}^{\frac{N}{2}}\left( \theta\right)=\sqrt{s!}\sqrt{\left(N-s\right)  !}\sqrt{n!}\sqrt{\left(N-n\right)  !}\\
&\times\sum_{v=\max\left[0,s-n \right] }^{\min\left[ N-n,s\right] }\frac{\left(-1 \right)^{v}\left[\cos\left(\theta \right)  \right]^{n+s-2v} \left[\sin\left(\theta \right)  \right]^{m-s+2v} }{v!\left(N-n-v \right)!\left(s-v \right)!\left(m-s+v \right)! }
\label{eq:refname1}
\end{align}

\begin{figure}[htbp]
\centering
\includegraphics[width=7cm]{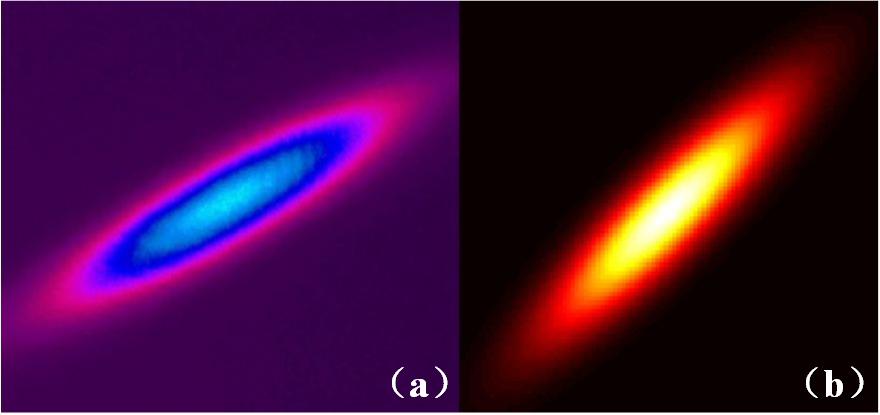}
\caption{The fundamental mode (a) in the experiment and (b) in the simulation.}
\label{fig:2}
\end{figure}

\begin{figure}[htbp]
\centering
\includegraphics[width=8cm]{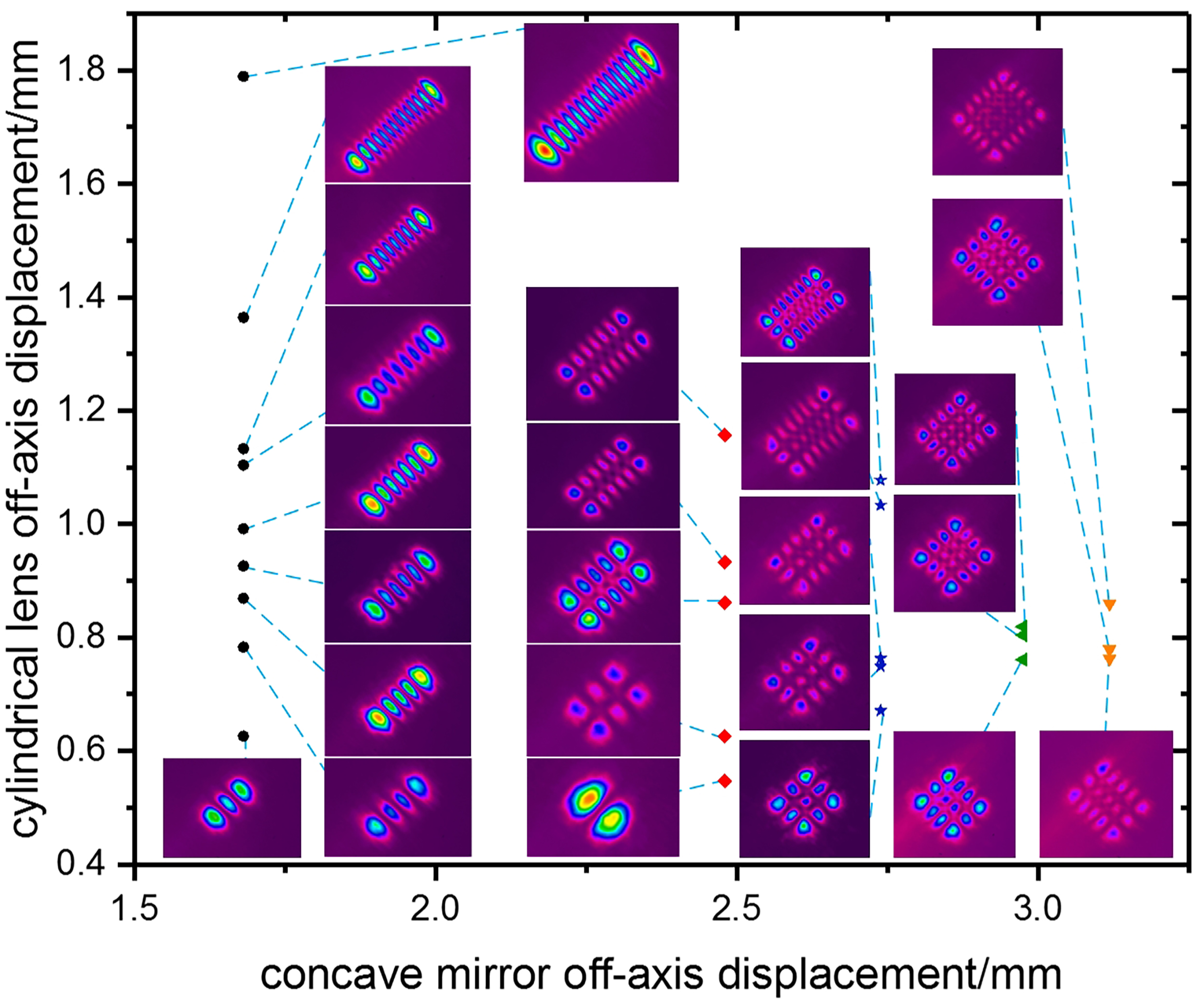}
\caption{Different generated HG modes with continually and independently tunable indices $\textit{m}$ and $\textit{n}$ with the off-axis of the cylindrical lens and the concave mirror.}
\label{fig:3}
\end{figure}

In the experiment, HG modes with continually and independently tunable indices $\textit{m}$ and $\textit{n}$ were generated with the off-axis of the cylindrical lens and R1. Direction of HG modes was limited by the setting angle of the intracavity cylindrical lenses. The order of HG modes in the dimension perpendicular to or along the generatrices of the cylindrical lenses changed with R1 off-axis along $y$-axis or one of the cylindrical lenses off-axis along $x$-axis, respectively. With all cavity elements coaxial, the astigmatic fundamental mode was shown as Fig.~\ref{fig:2} (a) at the pumping threshold of 0.56~W. Its simulation shown as Fig.~\ref{fig:2} (b), was consistent with the experimental result. While other elements remained coaxial, with the off-axis displacement of R1 along $y$-axis ($\Delta_{1}$) as 1.681~mm, the off-axis displacement of the first cylindrical lens along $x$-axis ($\Delta_{2}$) as 0.626~mm, 0.783~mm, 0.869~mm, 0.926~mm, 0.992~mm, 1.104~mm,1.133~mm, 1.364~mm, 1.789~mm, and pumping power 2.95~W, 3.37~W, 3.80~W, 4.01~W, 4.22~W, 4.43~W, 4.43~W, 5.77~W, 7.41~W, respectively, astigmatic HG modes HG$_{0,i}$ ($i$=2-8,13,15) were shown as Fig.~\ref{fig:3} the round dots row. With $\Delta_{1}$ as 2.480~mm, $\Delta_{2}$ as 0.547~mm, 0.626~mm, 0.861~mm, 0.933~mm, 1.157~mm, and pumping power 4.32~W, 4.97~W, 5.86~W, 7.14~W, 8.83~W, respectively, astigmatic HG modes HG$_{1,i}$ ($i$=0,2,4-6) were shown as Fig.~\ref{fig:3} the diamond dots row. With $\Delta_{1}$ as 2.739~mm, $\Delta_{2}$ as 0.671~mm, 0.749~mm, 0.763~mm, 1.033~mm, 1.077~mm, and pumping power 5.49~W, 6.06~W, 6.16~W, 7.93~W, 8.69~W, respectively, astigmatic HG modes HG$_{2,i}$ ($i$=2-4,6,7) were shown as Fig.~\ref{fig:3} the star dots row. With $\Delta_{1}$ as 2.975~mm, $\Delta_{2}$ as 0.761~mm, 0.804~mm, 0.819~mm, and pumping power 7.80~W, 8.06~W, 8.99~W, respectively, astigmatic HG modes HG$_{3,i}$ ($i$=2-4) were shown as Fig.~\ref{fig:3} the regular triangle dots row. With $\Delta_{1}$ as 3.119~mm, $\Delta_{2}$ as 0.763~mm, 0.779~mm, 0.860~mm, and pumping power 8.83~W, 8.83~W, 10.03~W, respectively, astigmatic HG modes HG$_{4,i}$ ($i$=2,3,5) were shown as Fig.~\ref{fig:3} the inverted triangle dots row. Modes like HG$_{0,i}$ ($i$=9-12,14), HG$_{1,i}$ ($i$=1,3), HG$_{2,5}$, and HG$_{4,4}$ were also generated, but not stable. In this system, HG modes could be controllably generated in two dimensions directly from the laser cavity. Due to the limitation of pumping power, the varying range of $\textit{m}$ was 0 to 4, and the difficulty of having large value of $\textit{n}$ grew as $\textit{m}$ increased. 
\begin{figure}[H]
\centering
\includegraphics[width=9cm]{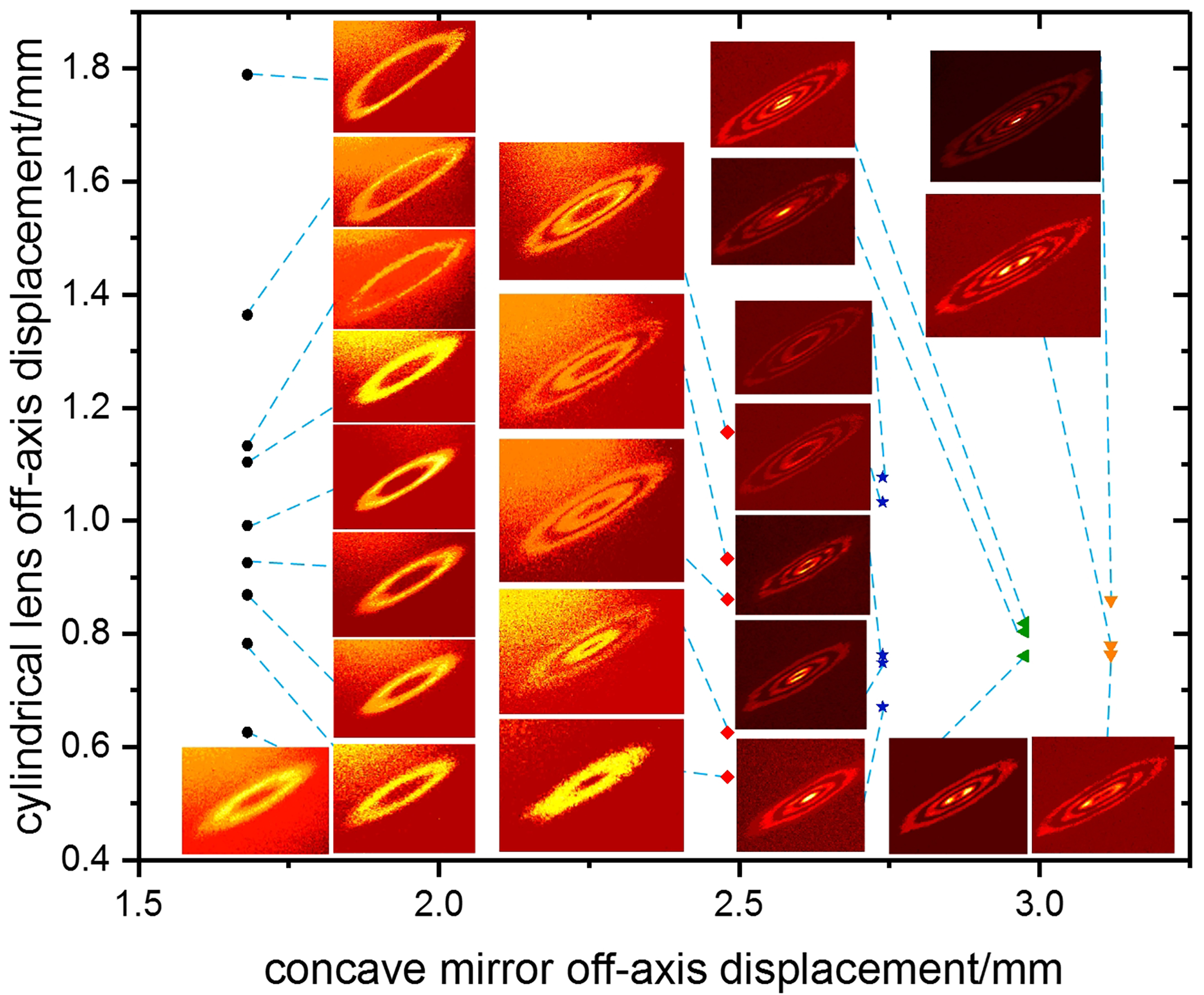}
\caption{Different OAM modes carrying continually and independently tunable azimuthal index $\ell$ and radial index $\textit{p}$ with the homologous off-axis displacements of the cylindrical lens and the concave mirror of HG modes.}
\label{fig:4}
\end{figure}
OAM beams with continually and independently tunable radial and azimuthal indices $(p,\ell)$ were generated after astigmatic HG modes going through the lenses group and the cylindrical lens outside the cavity. The HG$_{m,n}$ modes turned into OAM$_{p,\ell}$ modes. The radial index $\textit{p}$ was the smaller one of two parameters $\textit{m}$ and $\textit{n}$, and the azimuthal index $\ell$ was $\textit{n}$-$\textit{m}$. Matching HG modes in Fig.~\ref{fig:3}, OAM$_{0,i}$ ($i$=2-8,13,15) were shown as Fig.~\ref{fig:4} the round dots row, OAM$_{0,-1}$, OAM$_{1,i}$ ($i$=1,3-5) were shown as Fig.~\ref{fig:4} the diamond dots row, OAM$_{2,i}$ ($i$=0-2,4,5) were shown as Fig.~\ref{fig:4} the star dots row, OAM$_{2,-1}$, OAM$_{3,0}$, and OAM$_{3,1}$ were shown as Fig.~\ref{fig:4} the regular triangle dots row, and OAM$_{2,-2}$, OAM$_{3,-1}$, and OAM$_{4,1}$ were shown as Fig.~\ref{fig:4} the inverted triangle dots row, respectively. Limited by the generated HG modes, the tunable range of radial index $\textit{p}$ was from zero to four, and azimuthal index $\ell$ could be changed from -1 to 15, 0 to 5, -2 to 5, -1 to 1, 0 to 1, respectively.

\begin{figure}[H]
\centering
\includegraphics[width=8cm]{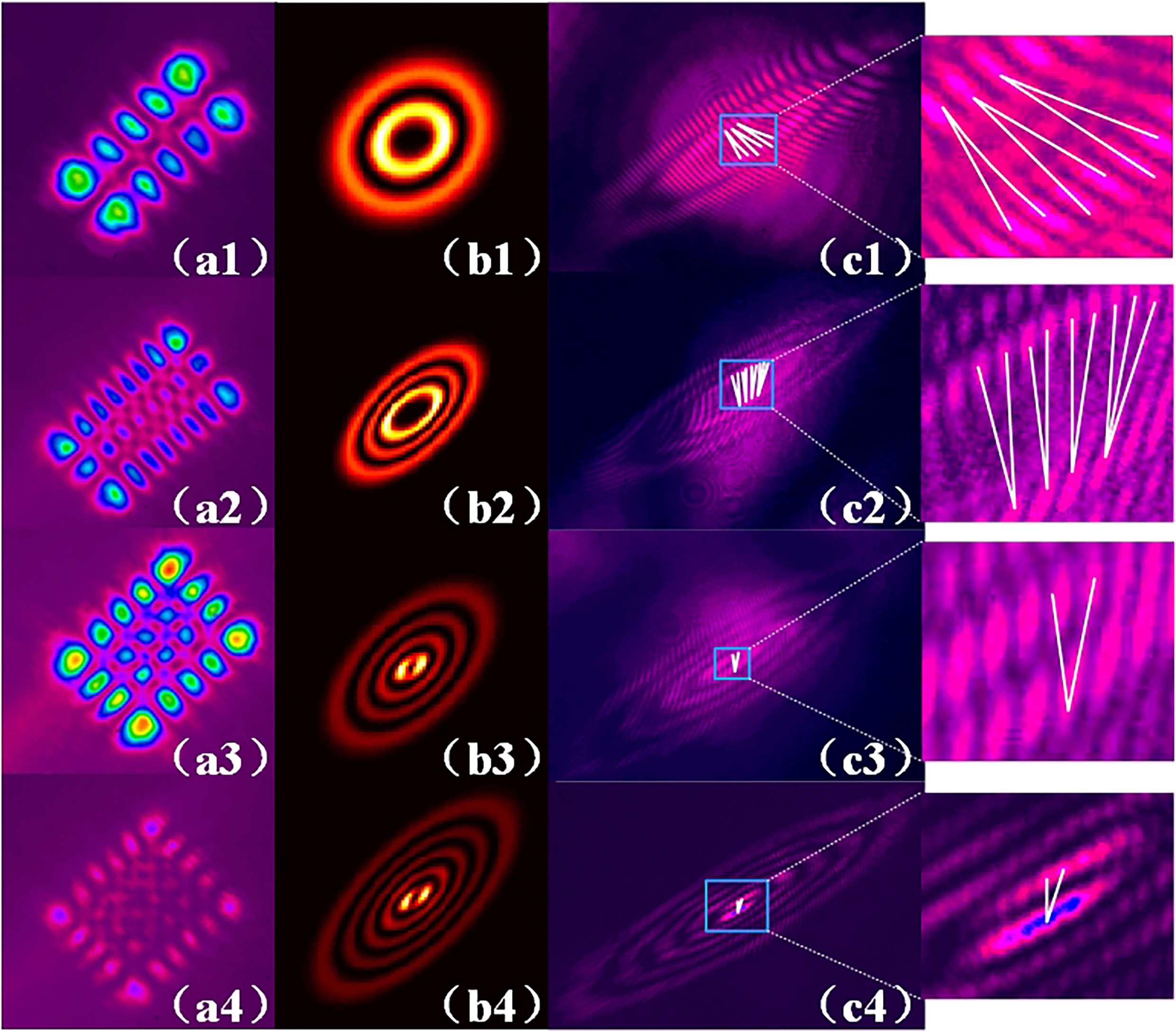}
\caption{ (a1-a4) The HG modes with (b1-b4) their homologous OAM beams in the simulation and (c1-c4) homologous interference patterns in the experiment.}
\label{fig:5}
\end{figure}

OAMs of the vortex beams were verified by interference patterns with the reference beam, as illustrated partially in Fig.~\ref{fig:5}. HG$_{1,4}$, HG$_{2,7}$, HG$_{3,4}$, HG$_{4,5}$ converted into OAM$_{1,3}$, OAM$_{2,5}$, OAM$_{3,1}$, OAM$_{4,1}$, and the interference patterns in Fig.~\ref{fig:5} (c1-c4) showed the topological charges three, five, one, one in each circle, respectively.
\begin{figure}[H]
\centering
\includegraphics[width=9cm]{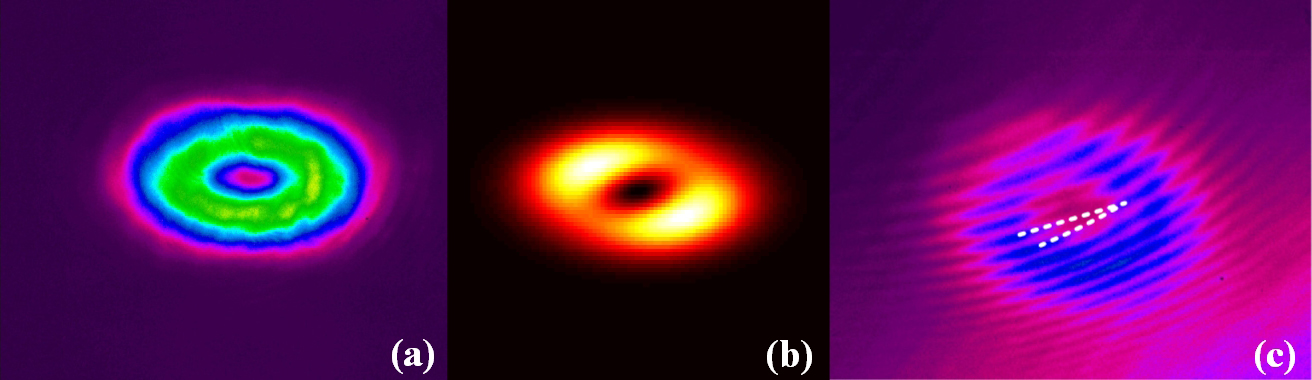}
\caption{$\pm1\hslash$ OAM modes generated directly from the cavity (a) in the experiment and (b) in the simulation and (c) the interference pattern in the experiment.}
\label{fig:6}
\end{figure}

As another important function of this cavity structure with the pair of cylindrical lenses set vertically in the cavity, OAM beams carrying $\pm1\hslash$ could be generated directly from the cavity. The pair of cylindrical lenses were set 21.6~mm away from Yb:CALGO and 35.7~mm apart from each other. $\Delta_{1}$ and $\Delta_{2}$ were controlled as 0.847~mm and 0.068~mm respectively at the pumping power 4.01~W. Based on the light path above, without the cylindrical lens outside the cavity, $\pm1\hslash$ OAM modes and its interference pattern were recorded by CCD1 shown in Fig.~\ref{fig:6} (a) and (c), respectively.

In conclusion, by putting a pair of cylindrical lenses into the cavity, controlling off-axis displacements of the concave mirror R1 along $y$-axis and the first cylindrical lens along $x$-axis, the indices of HG$_{m,n}$ modes were tuned continually and independently in two dimensions for the first time. With the value of $\textit{m}$ as 0 to 4 limited by the pumping power, the tunable range of $\textit{n}$ was 0 to 15, 0 to 6, 2 to 7, 2 to 4, and 2 to 5, respectively. After astigmatic conversion through the lenses group and the cylindrical lens outside the cavity, HG modes changed into OAM beams carrying continually and independently tunable radial and azimuthal indices $(p,\ell)$. With the value of $\textit{p}$ as 0 to 4, azimuthal index $\ell$ could be tuned from $-1$ to 15, 0 to 5, $-2$ to 5, $-1$ to 1, and 0 to 1, respectively. As another important function of this cavity structure, with certain parameter control, $\pm1\hbar$ OAM modes could be generated directly from the cavity. All OAMs mentioned above were verified by the interference patterns, and beams carrying OAMs in the experiments had great consistency with those in the simulations. This work has great potential in increasing structured light tunability at the source in a simple and cost-saving way, and enriching the related applications.

\medskip

\noindent\textbf{Fundings.} National Natural Science Foundation of China (61875100).


\bibliography{sample}






\end{document}